\documentclass[a4paper]{llncs}

\usepackage[pdftex]{graphicx} 
\usepackage{array}
\usepackage[caption=false,font=footnotesize]{subfig}

\usepackage{balance}		
\usepackage[utf8]{inputenc}
\usepackage{booktabs}
\usepackage{multirow}
\usepackage{pifont} 
\usepackage{url}
\usepackage{hyperref}

\usepackage{listings}
\newcommand{\code}[1]{\lstinline|#1|}

\usepackage[warn]{textcomp}

\newif\ifremarks
\remarksfalse 

\begin{document}

\title{AntibIoTic: Protecting IoT Devices Against DDoS Attacks}

\author{
Michele De Donno\inst{1}
Nicola Dragoni\inst{1,2}
Alberto Giaretta\inst{2} and
Manuel Mazzara\inst{3}
\institute{DTU Compute, Technical University of Denmark, Denmark
\and Centre for Applied Autonomous Sensor Systems, \"{O}rebro University, Sweden
\and Innopolis University, Russian Federation}}


\maketitle

\begin{abstract}
The 2016 is remembered as the year that showed to the world how dangerous 
Distributed Denial of Service attacks can be. Gauge of the disruptiveness of DDoS attacks is 
 the number of bots involved: the bigger the botnet, the 
more powerful the attack. This character, along with the increasing 
availability of connected and insecure IoT devices, makes DDoS and IoT the 
perfect pair for the malware industry. In this paper we present the main idea behind AntibIoTic, a palliative solution to prevent DDoS attacks perpetrated through IoT devices. 
\end{abstract}

\section{The AntibIoTic Against DDoS Attacks}
Today, it's a matter of fact that IoT devices are extremely poorly secured and 
many different IoT malwares are exploiting this insecurity trend to spread 
globally in the IoT world and build large-scale botnets later used for 
extremely powerful cyber-attacks \cite{York_2016_Dyn,Hilton_2016_Dyn}, especially Distributed Denial of Service (DDoS) \cite{INSERT17}. Therefore, the main problem 
that has to be solved is the low security level of the IoT cosmos, and that is where AntibIoTic comes in.

What drove us in the design of AntibIoTic is the belief 
that the intrinsic weakness of IoT devices might be seen as the solution of the 
problem instead of as the problem itself. In fact, the idea is to use
the vulnerability of IoT units as a means to grant their security: like an 
antibiotic that enters in the bloodstream and travels through human body 
killing bacteria without damaging human cells, 
AntibIoTic is a worm that infects vulnerable devices and creates 
a white botnet of safe systems, removing them from the clutches of other 
potential dangerous malwares. Basically, it exploits the most efficient 
spreading capabilities of existing IoT malwares (such as Mirai) in order to 
compete with them in exploiting and infecting weak IoT hosts but, once control 
is gained, instead of taking advantage of them, it performs several 
operations aimed to notify the owner about the security threats of his device 
and potentially acting on his behalf to fix them.
In our plans, AntibIoTic will raise the IoT environment 
to a safer level, making the life way harsher for DDoS capable
IoT malwares that should eventually slowly disappear.
Moreover, the whole solution has been designed including some functionalities 
aimed at creating a bridge between security experts, devices manufacturers and 
users, in order to increase the awareness about the IoT security problem and 
potentially pushing all of them to do their duties for a more secure global Internet.

Similar approaches have been occasionally tried so 
far~\cite{LinuxWifatch_2015_Symantec,Hajime_2017_Symantec,BrickerBot_2017_BleepingComputer} but, to the best of our knowledge, they have mostly been rudimentary and not
documented pieces of code referable to crackers (or, as wrongly but
commonly named, hackers) that want to solve the IoT security problem by taking 
the law into their own hands, thus poorness
or even lack of preventive design and documentation are the common traits. 
Nevertheless, these attempts are the proof that the proposed solution is 
feasible and parts of their source code have been published under OpenGL 
license~\cite{LinuxWifatch_GitHub_2015}, which makes them 
reusable for the implementation of AntibIoTic.

The  paper continues presenting a high level overview of the 
AntibIoTic functionalities and infrastructure, respectively in Sections~\ref{sec:functionalities}-\ref{sec:infrastructure}. Then, a comparison with existing similar approaches is given in Section~\ref{sec:Twins}, and legal and ethical implications are discussed in Section~\ref{sec:EthicalLegalImplications}.

\section{AntibIoTic Functionalities}
\label{sec:functionalities}

Looking from an high level perspective, the AntibIoTic 
functionalities include, but are not limited to:
\begin{itemize}
	\item \textit{Publish useful data and statistics} - Thanks to the 
	infrastructure behind the AntibIoTic worm, IoT security best 
	practises and botnet statistics computed from the data collected by the 
	worm, can be published online and made available to 
	anyone interested (not only experts);
	\item \textit{Expose interactive interfaces} - Interactive interfaces 
	with different privileges are also
	publicly exposed in order to let anyone join and improve the 
	AntibIoTic 
	solution;
	\item \textit{Sanitize infected devices} - Once the control of a weak 
	device is gained, the AntibIoTic worm cleans it up from other 
	possibly running malicious malwares and secure its perimeter avoiding 
	further 
	intrusions;
	\item \textit{Notify device owners} - After making sure the device has 
	been sanitized, the AntibIoTic worm tries to notify the device 
	owner pointing out the device vulnerabilities. The notification aim is to 
	make the owner aware of the security threats of his device and give him 
	some advices to solve them;
	
	\item \textit{Secure vulnerable devices} - Once notified the device owner, 
	if the security threats haven't been fixed yet,	the AntibIoTic 
	worm starts to apply all the possible security best practises aimed to 
	secure the device. For instance, it may change the admin credentials and 
	update the firmware;

	\item \textit{Resistance to reboot} - AntibIoTic incorporates a 
	basic mechanism that let it keep track of all spotted vulnerable devices 
	and, 
	if a target device reboot occurs, it is able to reinfect them as soon as 
	they 
	are 
	up and running. Moreover, in order to avoid the worm to be wiped off from 
	device memory by a simple reboot, the AntibIoTic worm may also use 
	an advanced mechanism to persistently settle into the target system by 
	modifying its startup 
	settings.
\end{itemize}

Please consider that the functionalities presented above are only an high level 
summary of the AntibIoTic set of functions, aimed to give the reader a 
first conception of the solution. A more clear explanation of the 
AntibIoTic modus operandi is given in 
Section~\ref{sec:AntibioticDetails}.

\subsection{Real World Scenarios}
Given the basic idea behind AntibIoTic, in this subsection we will get 
through some different working scenarios that the AntibIoTic worm 
could face during its propagation and in which a subset of the aforementioned 
functionalities are used. Each scenario will be presented using an high level 
graphical workflow and a brief textual explanation.


\subsubsection{Scenario 1 - \textit{Awareness notification}}
\begin{figure}[t]
	\centering
	\includegraphics[clip=true, width=0.6\textwidth, trim=4cm 9.3cm 4.2cm
	0.2cm]{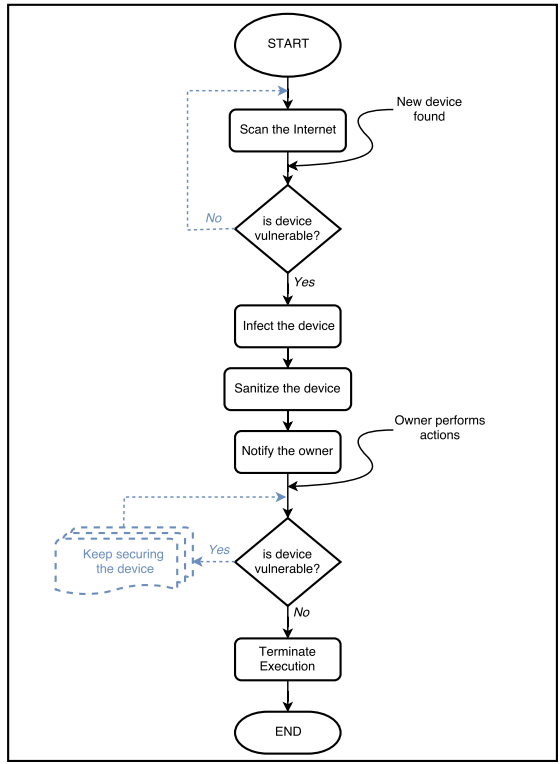}
	\caption{Device owner secures its device after 
		receiving the AntibIoTic notification 
		\label{fig:Scenario1}}
\end{figure}
The first scenario is the one shown in 
Figure~\ref{fig:Scenario1}. It is the ideal situation in 
which as soon as the device owner sees the AntibIoTic notification, he 
performs 
some of the suggested operations in order to secure the device.

First of all, AntibIoTic scans the Internet looking for IoT weak 
devices. As soon as a vulnerable device is found, it is infected and sanitized 
in order to secure its perimeter and ensure that no other malwares are 
in execution on the same device. Subsequently, the awareness notification is 
sent to the owner pointing out the security threats of the device and some 
possible countermeasures to solve them. Then, the scrupulous device owner looks 
at the notification and secures its device following the guidelines given by 
AntibIoTic. At this point, the IoT device is not vulnerable anymore 
thus the AntibIoTic intent has been reached and it can terminate its 
execution freeing the device.
More elaborate (and, probably, real) cases, in which the owner doesn't perform 
any action to increase the security level of its device, are presented in the 
following scenarios.


\subsubsection{Scenario 2 - \textit{Credentials change on a rebooted device}}
\begin{figure}[t]
	\centering
	\includegraphics[clip=true, width=0.8\textwidth, trim=0.2cm 7.3cm 0.2cm 2.8cm]{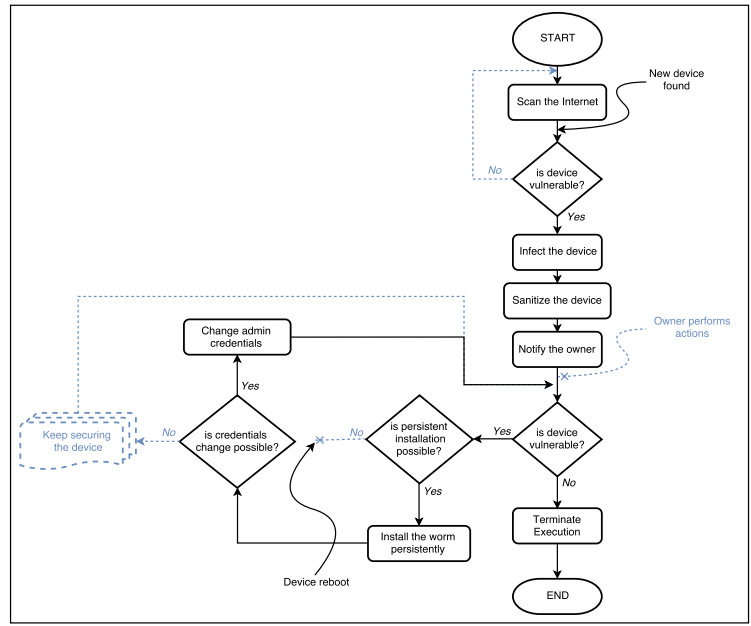}
	\caption{Credentials change after persistent installation
		\label{fig:Scenario2}}
\end{figure}

The second scenario is depicted in 
Figure~\ref{fig:Scenario2}. In this case, the device owner is 
impassive to the AntibIoTic notification and a device reboot occurs 
while AntibIoTic is performing its security tasks. However, thanks to 
the persistent installation and the credentials change functionalities, 
AntibIoTic is able to secure the device as well.

As seen in the first scenario, at first AntibIoTic looks for a 
vulnerable 
device, infects and sanitizes it, and notifies its owner. Nevertheless, in this 
case, the device owner either ignore or doesn't see the AntibIoTic 
notification, thus he performs no actions. Whereby, AntibIoTic 
starts to secure the device by checking if it's possible to settle down on the 
hosting device in order to resist to potential reboots. In this scenario, we 
are hypothesizing that the persistent installation is possible hence the 
AntibIoTic worm persistently settles down on the vulnerable device.
Now, let's suppose a device reboot occurs. However, since AntibIoTic 
has been 
persistently installed on the device, after the reboot it starts again and 
quietly picks its tasks up where it left off. It checks if a 
credentials change is possible. In this scenario, we are supposing that it is 
allowed, thus the AntibIoTic worm changes the admin credentials. Now, 
thanks to the security actions performed, the target device is 
not vulnerable anymore, hence the AntibIoTic worm terminates its 
execution and frees the device.

\subsubsection{Scenario 3 - \textit{Firmware update of a reinfected device}}
\begin{figure}[t]
	\centering
	\includegraphics[clip=true, width=0.8\textwidth, trim=0.1cm 8.3cm 0.2cm 0.5cm]{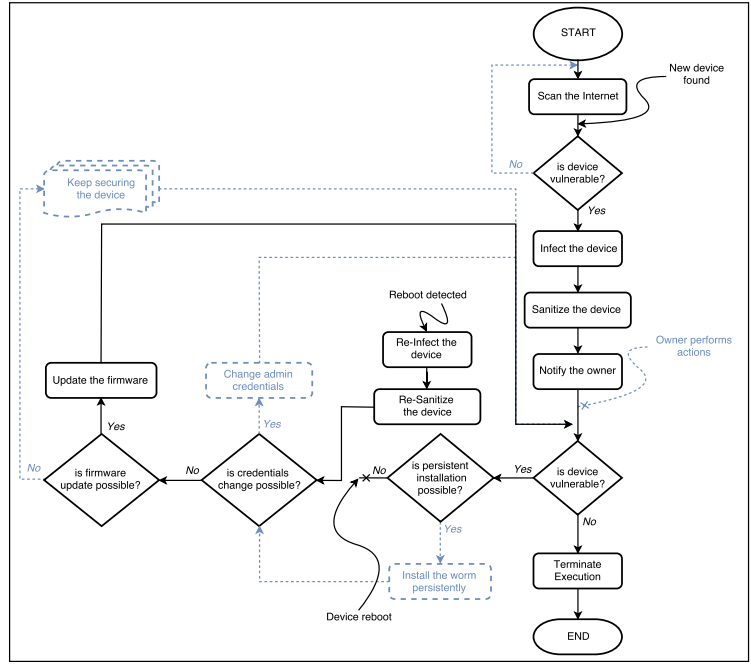}
	\caption{Firmware update after reinfection
		\label{fig:Scenario3}}
\end{figure}

The third scenario is shown in Figure~\ref{fig:Scenario3}. It is 
a harsh environment for AntibIoTic, since persistent installation and 
credentials change are not possible and a device reboot occurs while it is 
performing its duties. Nevertheless, thanks to its reboot-resistant design, it 
is able to reinfect the device and secure it through a firmware update.	

The first part of the workflow moves along same lines as the aforementioned 
scenarios: AntibIoTic finds a vulnerable device, infects and sanitizes 
it, notifies the owner. Also in this case the owner doesn't perform any action, 
so the AntibIoTic worm checks if the persistent 
installation is possible. In this case, we are hypothesizing that it is not 
allowed and that a device reboot occurs before AntibIoTic can perform 
any other operation. So, the hosting device is rebooted and our worm is wiped 
off from its memory. Nevertheless, the AntibIoTic infrastructure 
detects the 
reboot and monitors the target device to reveal whenever it is up and running 
again. As soon as again available, the vulnerable device is reinfected and 
resanitized by the AntibIoTic worm. Now, it continues to perform its 
actions checking if credentials change is possible. We are supposing that it is 
not, so AntibIoTic looks if a firmware update is feasible. Let's 
suppose that it is and our worm downloads and installs an up-to-date firmware 
on the hosting device. Now, the target device is safe and the 
AntibIoTic worm can stop its execution freeing the device.

\section{Overview of AntibIoTic Infrastructure}\label{sec:AntibioticDetails}
\label{sec:infrastructure}

\begin{figure}[!h]
	\centering
	\includegraphics[clip=true, width=0.7\textwidth, trim=1.8cm 4.4cm 1.7cm
	0.8cm]{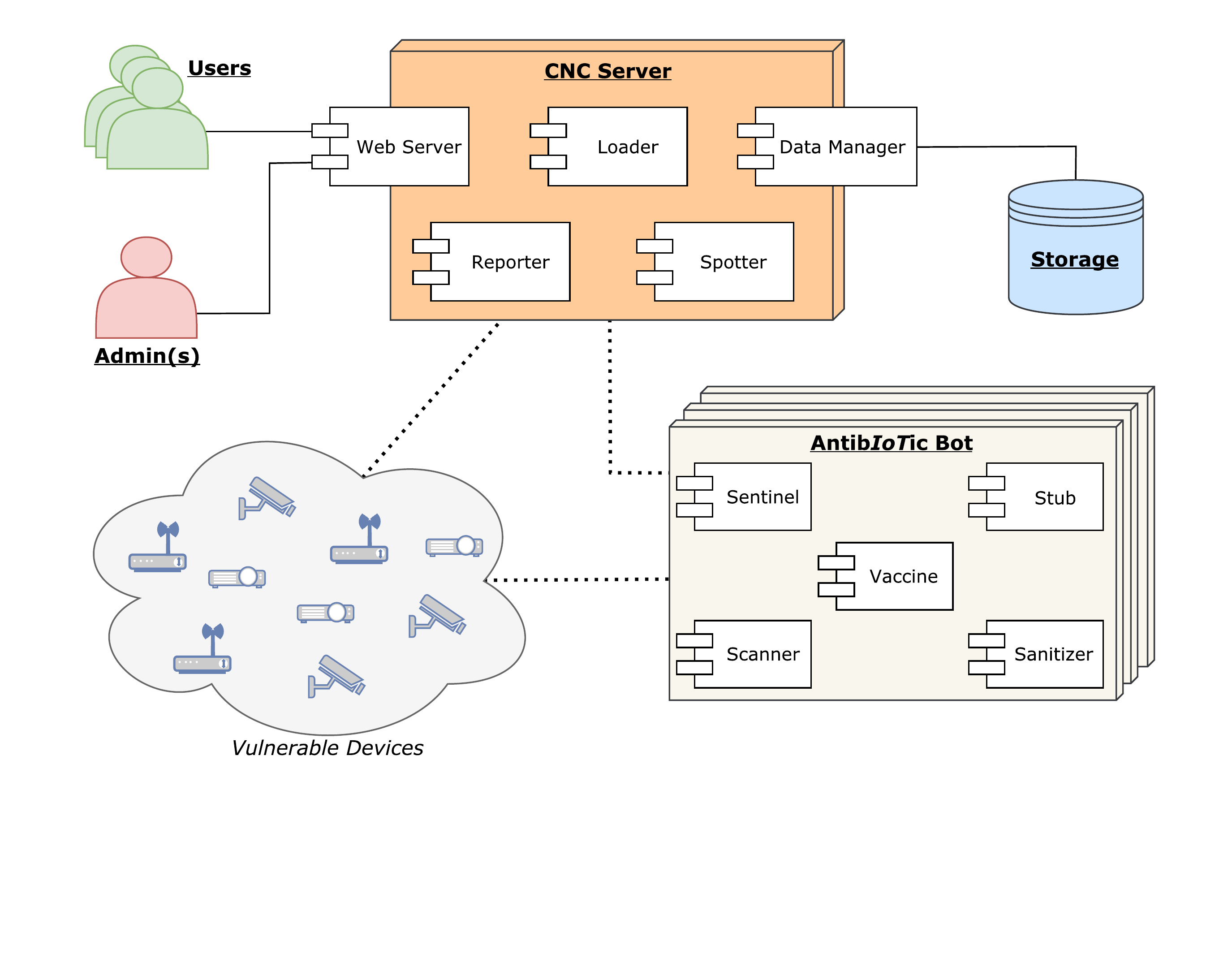}
	\caption{AntibIoTic 
		infrastructure\label{fig:AntibioticInfrastructure}}
\end{figure}

The overall architecture of AntibIoTic 
(Figure~\ref{fig:AntibioticInfrastructure}) is mostly
arisen from the Mirai infrastructure. This choice has been driven by the
strong evidence of robustness and efficiency that Mirai gave to the
world the last year as well as by the ascertainment that, despite its
efficiency, the Mirai architecture is relatively simple and most of the source
code needed for its implementation is already available online 
\cite{MiraiSourceCode_GitHub_2016}, which makes it easily reusable.

At a macroscopic level, the AntibIoTic infrastructure is made of 
several components and actors interacting with each other.

\subsection{Command-and-Control (CNC) Server} 
It is the central 
component of the infrastructure. It is in charge of performing several 
tasks interacting with other actors and components. It is composed of 
different modules:
	\begin{itemize}
		\item \textit{Web Server} - 
		It is the module that exposes the botnet 
		human interface with human actors. It shows 
		some useful data and live statistics and supports the interaction with 
		two type of actors, 
		each allowed to perform different 
		operations: \textit{user}, \textit{admin};
		\item \textit{Reporter} - 
		It is the module in charge of 
		receiving and processing vulnerability results and relevant 
		notifications sent by AntibIoTic Bots;
		\item \textit{Spotter} -	
		It is the module that handles the keep-alive messages continuously sent 
		from AntibIoTic Bot Sentinel modules, ensuring a working 
		connectivity with 
		each infected devices. If for some reason (e.g., device 
		reboot) the communication between the Spotter and the device is lost, 
		the former immediately notifies the Loader to periodically 
		try to gain	the control of the insecure device again;
		\item \textit{Loader} - 
		It is the module that uses the received vulnerability 
		results to remotely infect and gain control of insecure devices. It is 
		also in 
		charge of loading up-to-date modules on and sending commands to 
		AntibIoTic Bots;
		\item \textit{Data Manager} - 
		It is the module which exposes the API to access all data saved on 
		the Storage. Each module of the CNC Server interacts with Data Manager 
		to 
		perform any operation to local data.
	\end{itemize}
	All data and files relevant for the whole infrastructure are saved in the 
	\textbf{Storage}. It is accessible by all the modules of the CNC Server 
	through the Data Manager.	
	
\subsection{AntibIoTic Bot} 
It is the component running on
vulnerable devices with the aim of securing them. It is composed of distinct
modules in order to perform different tasks:
	\begin{itemize}
		\item \textit{Stub} -
		It is the main module of the worm. It is in charge of starting most of 
		the 
		other modules and listening for further commands or module updates 
		received from the Loader module of the CNC Server;
		\item \textit{Sentinel} - 
		It is the module in charge of continuously communicating with
		the Spotter module of the CNC Server. It mainly sends keep-alive 
		messages 
		or local reboot notifications to the Spotter;
		\item \textit{Scanner} -
		It is the module that scans for new vulnerable IoT devices using a list 
		of well-know credentials.
		Once a weak device is found, its information are sent back to the 
		Reporter module of the CNC Server. This module corresponds to the Mirai 
		Bot Scanner module;
		\item \textit{Sanitizer} - 
		It is the module that cleans up the target device by both eradicating 
		other potential running malwares and performing safety operations aimed 
		to secure the device from further intrusions. This module is alike the 
		Mirai Bot Killer module;
		\item \textit{Vaccine} - 
		It is the module that performs several operations directed to increase 
		the security level of the target device. The number and type of 
		performed actions depend on the nature of the hosting device and some 
		of them 
		can involve human interaction.
	\end{itemize}
	
\subsection{Users and Admin} 
Users are one of the human actors involved in the 
AntibIoTic infrastructure. It can interact with the Web Server 
module of the CNC Server 
just to get known about relevant data and live statistics or it can 
actively 
contribute to the project by submitting new information about additional 
security threats 
affecting IoT devices.

Finally, Admin is the human actor in charge of supervising the 
AntibIoTic infrastructure. It can perform operations on data 
saved in the Storage as well as send control commands to the botnet 
(further details and consideration about this last option will follow). It 
is also in charge of reviewing information submitted by users in order to discard them 
or accept them and accordingly update the involved AntibIoTic 
modules.	

\section{AntibIoTic and Its "Twins"}
\label{sec:Twins}

As previously mentioned, there are already some so-called 
"vigilantes"~\cite{LinuxWifatch_2015_Symantec,Hajime_2017_Symantec,BrickerBot_2017_BleepingComputer} out there which 
have been built with an aim similar to the AntibIoTic one, thus it is 
more than legitimate to wonder:
"why is AntibIoTic better than its twins?". We won't directly answer to
the question, but we want to address it by providing a comparison between 
AntibIoTic and the other existing solutions (also referred as 
"twins"), which is summarized in Table~\ref{tab:AntibIoTicTwins}.

\begin{table*}[hbt]
	\centering	
	\caption{Comparison between AntibIoTic and similar	solutions}
	\label{tab:AntibIoTicTwins}
	\resizebox{0.9\textwidth}{!}{%
		\begin{tabular}{>{\raggedright\arraybackslash}p{4.7cm} c c c c} 
			\toprule
			& \multicolumn{3}{c}{\textbf{Twins}} & 
			\multirow{2}{*}{\textbf{AntibIoTic}}\\ 
			\cmidrule(lr){2-4}
			& BrickerBot & Hajime & Linux.Wifatch \\
			\midrule
			Publicly documented & - & - & - & \ding{51} \\			
			Create awareness and encourage synergy & 
			- & - & \ding{51} & \ding{51}\\			
			Notify infected device owners & - & \ding{51} & 
			\ding{51} & \ding{51}\\ 		
			Temporary security operations & 
			\ding{51} & \ding{51} & 
			\ding{51} & \ding{51}\\ 
			Permanent security operations & - & - & - &\ding{51}\\ 
			\bottomrule
		\end{tabular}%
	}
\end{table*}

First of all, we do not claim that our solution is absolutely better than 
the others, basically because we have not enough data to assert 
it. Indeed, to the best of our knowledge, the 
existent solutions are not documented at all and the only sources of 
information that we can use to make a comparison are some security analysis and 
reverse engineering works found online, which try to point out the main traits 
of each white worm.  The closest thing to a documentation that we saw in the 
wild is 
the Linux.Wifatch GitHub repository~\cite{LinuxWifatch_GitHub_2015} which 
provides 
a rough explanation of the source code folders hierarchy and some general 
comments about the authors' purpose. Nevertheless, it doesn't give a clear 
presentation of the whole infrastructure and it doesn't explain how each 
component interacts with 
the others, thus we won't consider it as an actual documentation.
That is, for us, the first  
plus point for AntibIoTic, since with this work we are providing a  
presentation as clear as possible of our solution that can be intended as 
documentation. Let's now proceed toward an high level functional analysis in 
order to continue the comparison.

Starting the functionalities review from the AntibIoTic 
infrastructure, it soon becomes evident the bridge that the CNC Server wants to 
create between AntibIoTic and the people. Indeed, our 
solution wishes to interact with experts, devices manufacturers and common 
users in order to show them how critique and dangerous the current IoT security 
situation is and potentially pushing them to do their best (e.g., put into 
practice the basic security recommendation) to  improve it. 
Moreover, AntibIoTic give them the chance of interacting with the 
whole infrastructure by submitting useful information that could be used by the 
white worm to be more powerful and effective. That is because our aim is 
not to build a sneaky worm that stabs the device owners in the back and which 
the people should be scared of, but we want to build a white worm that owners 
are happy to see on their devices since it helps them by giving some advices 
or by securing the devices in their behalf. Apparently, no one of the 
AntibIoTic twins tries to create the same empathy with the common 
people but Linux.Wifatch, whose authors published the source code and explained 
their purpose encouraging people to take part in the project. Therefore, even 
if the way in which it is performed is different from the AntibIoTic 
approach, we can say that also Linux.Wifatch is aimed to both create awareness 
about the IoT security problem and encourage the collaboration of people to 
implement a white 
worm that tries to improve the current situation.

Talking about the actual worm functionalities, that is where 
most of the similarities are. 
First of all, almost all the twins notify the 
infected IoT device owner telling him that his device is insecure and some 
security operations are needed. That is, more or less, the same behaviour of 
AntibIoTic. Secondly, all the twins try to perform some 
operations aimed to secure the target device. The type of 
performed operations differs from solution to solution and from hosting device 
to hosting device but the high level result is almost always the same: keep the 
device safe until the memory is wiped off. 
The same goal is reached by AntibIoTic but, unlike its twins, it goes 
ahead and tries to permanently secure the hosting device. The only twin that 
tries 
to accomplish the same goal is BrickerBot. However, relevant is to point out 
the way in which 
BrickerBot achieves its aim. It usually tries to permanently secure the 
hosting unit without damaging it but, if that is not possible, it writes random 
bits on the device storage often bricking it and requiring the owner to replace 
it. 
This kind of malicious behaviour has been classified as a \textit{Permanent 
Denial of 
Service} (PDoS) attack~\cite{ERT_2017_Radware} and we strongly disapprove of 
it, because it does not fit the "white" purpose 
of this class of worms. So, even if the aim of BrickerBot author is to 
permanently secure IoT devices~\cite{BrickerBot1_2017_BleepingComputer}, and 
somehow it actually achieves it (insecure devices are irredeemably 
damaged, thus put offline), in our comparison we will not consider BrickerBot 
as a white worm that permanently secure IoT devices because the way in which 
it is done can not be treated as legitimate and thus accepted.

To sum up, from the Table~\ref{tab:AntibIoTicTwins} the main threads of the 
comparison between AntibIoTic and the other similar solutions can be 
extrapolated. All the existing solutions basically lack of a solid 
documentation that clarifies their aim and structure. Moreover, even if most of 
them notify the owner of the infected device and push him to secure it, they do 
not try to create a connection with all people in order to increase the global
awareness about the IoT security problem and stimulate a profitable interaction 
with them to improve the situation. 
Furthermore, as widely said by several security experts, 
the main problem of all the AntibIoTic twins is that they usually have 
a short lifespan on the target device since their actions are only temporary 
and,
as soon as the hosting device is rebooted, they are wiped off from 
memory and the unit goes back to its unsafe state. That is not applicable to 
AntibIoTic, since it is provided with some unique and smart 
functionalities, such as resistance to reboot and firmware update, that allow 
it to resist to reboot and permanently secure infected devices. 

Basically, 
AntibIoTic can be considered an 
evolution of the current white worms which picks the best from them and also 
adds some new functionalities to both fix their mistakes and propose a new idea of joint participation to the IoT security process.

\section{Ethical and Legal Implications}
\label{sec:EthicalLegalImplications}

It is undeniable that the proposed solution drags on some ethical and legal 
implications, mainly arisen by the intent of gaining control of unaware 
vulnerable devices, even if it is done for security purposes.

Sometimes the line between ethical and unethical behaviour is a fine one and,
whenever we try to design a possible solution to a malicious conduct, we can 
not be exempt from asking ourselves if our proposal goes too far. Even though 
AntibIoTic is motivated by the desire of fixing a harsh situation 
created by firms unforgivable negligence, it requires to break-in third-party 
devices without the owners' explicit consent, which is an illegal 
and prosecutable practice in several countries. Nevertheless, we can not ignore 
that, accordingly to various legislations, also the very action of failing to 
protect your own device and unwillingly participating to a malicious action 
could be considered illegal. This entails that our solution could be warmly 
welcomed and tolerated by the less knowledgeable users worried to incur in 
possible prosecution, but unable to apply themselves a more adequate and 
stronger security policy.

Somehow, we can think about AntibIoTic as a scapegoat that secures 
IoT devices and impedes them to cause any harm. A scapegoat that accepts the 
risk to be accused for the hosts infection, but both increases the IoT security 
and keeps safe the users avoiding them to incur into tough prosecutions.

Therefore, what we are indirectly asking to the users is to blindly trust that 
both AntibIoTic and its maintainers are well-meaning. We known that it 
is a greedy claim, but we also believe that it can be achieved through the 
power of a large community that supports and trusts the project, and which is 
willing to work in order to improve it. Accordingly, what we are basically 
thinking of, is a single word: \textit{open-source}. 
We strongly feel, to such an extent that we would define it 
mandatory, that AntibIoTic, as well as other similar approaches, 
should be released as open-source projects in order to fulfil two main 
benefits.

The first one is to build trust between the project and the IoT users, 
because only a strong trust into the project solidity and well-meaning can 
ensure the people collaboration.
Furthermore, we highlight that the more discretion is left to 
AntibIoTic 
admins, the more concerns will be risen into the device owners when it is 
asked 
them to trust a stranger to fully control their device. That is why, even if 
the AntibIoTic capabilities are completely transparent, 
the discretion power granted to the admins should be as limited as 
possible, ideally giving them only the option to shut down the whole botnet 
or release a single device, if required.

However, supposing for a moment that a high level of trust can be reached, we 
do not pretend to be considered better than others, hence we know that the 
resulting white botnet could always being hacked and used for malicious 
purposes. That is where the second open-source benefit comes in: 
an open-source project would attract other white-hat volunteers and companies 
that share our willingness to fight the IoT security threats, which would 
ensure a more updated, efficient and reliable software.

Truth be told, we are very concerned about users' privacy and we feel that the 
path traced by AntibIoTic should not be taken by anyone, because it 
could unexpectedly backfire and expose the vulnerabilities to malicious users, 
no matter if criminal organisations or intelligence agencies, that could 
exfiltrate highly-sensitive personal data.
The only reason why we suggest this solution, continuously stressing about the 
transparency requirements, is that the current situation is beyond any control 
and something has to be done before it gets even worse.

We are basically in front of the eternal dispute between freedom 
and security, and we are aware that the very right answer does not exist. 
However, to conclude, since we strongly believe that "my 
freedom ends where yours begins", we would like to leave the reader with a 
final question: \textit{what should we do when your freedom affects our 
security?}


\section{Conclusion}
\label{sec:conclusion}
In this paper we have presented the main idea behind AntibIoTic, a system to prevent DDoS attacks perpetrated through IoT devices. The functionalities of the system have been listed and some scenarios discussed. Comparison with similar approaches provides evidence that AntibIoTic represents a promising solution to the DDoS attacks problem in the IoT context. 

The key task of future work consists in the full implementation and evaluation of the system. In particular, architectural design has to be considered (or reconsidered) thoroughly. The architecture described in Figure~\ref{fig:AntibioticInfrastructure} shows a number of interacting components that need to scale up as the number of devices also scale up. It has has been shown that scalability issues can naturally be solved by use of microservice architecture~\cite{Dragoni2017,DLLMMS2017}, and that large-size companies have already implemented migrations to this architectural style~\cite{DDLM2017}. Furthermore, specific programming languages are available to support microservice architecture~\cite{Guidi2017,Safina2016}. Full deployment of the system should consider a migration to microservice, possibly making use of a suitable language and relying on the expertise of our team on the matter. Finally, a project on microservice-based IoT for smart buildings is currently running~\cite{Salikhov2016a,Salikhov2016b}, and it certainly represents a solid case study for experimentation and validation.

\bibliographystyle{ieeetr}
\bibliography{ms}

\end{document}